\documentclass[aps,a4paper,superscriptaddress,showpacs,preprintnumbers,amsmath,amssymb]{revtex4}

\usepackage{ulem}

\usepackage{psfrag} \usepackage{graphicx} \usepackage{dcolumn}
\usepackage{color} \usepackage{latexsym,amsfonts} \usepackage{bm}
\usepackage{amssymb}
\begin{document}

\title{Precision Analysis\\ of Electron Energy Spectrum and Angular
  Distribution\\ of Neutron $\beta^-$--Decay with Polarized Neutron
  and Electron}

\author{A. N. Ivanov}\email{ivanov@kph.tuwien.ac.at}
\affiliation{Atominstitut, Technische Universit\"at Wien, Stadionallee
  2, A-1020 Wien, Austria}
\author{R.~H\"ollwieser}\email{roman.hoellwieser@gmail.com}\affiliation{Atominstitut,
  Technische Universit\"at Wien, Stadionallee 2, A-1020 Wien,
  Austria}\affiliation{Department of Physics, New Mexico State
  University, Las Cruces, New Mexico 88003, USA}
\author{N. I. Troitskaya}\email{natroitskaya@yandex.ru}
\affiliation{Atominstitut, Technische Universit\"at Wien, Stadionallee
  2, A-1020 Wien, Austria}
\author{M. Wellenzohn}\email{max.wellenzohn@gmail.com}
\affiliation{Atominstitut, Technische Universit\"at Wien, Stadionallee
  2, A-1020 Wien, Austria} \affiliation{FH Campus Wien, University of
  Applied Sciences, Favoritenstra\ss e 226, 1100 Wien, Austria}
\author{Ya. A. Berdnikov}\email{berdnikov@spbstu.ru}\affiliation{Peter
  the Great St. Petersburg Polytechnic University, Polytechnicheskaya
  29, 195251, Russian Federation}

\date{\today}

\begin{abstract}
We give a precision analysis of the correlation coefficients of the
electron--energy spectrum and angular distribution of the
$\beta^-$--decay and radiative $\beta^-$--decay of the neutron with
polarized neutron and electron to order $10^{-3}$. The calculation of
correlation coefficients is carried out within the Standard model with
contributions of order $10^{-3}$, caused by the weak magnetism
and proton recoil, taken to next--to--leading order in the large
proton mass expansion, and with radiative corrections of order
$\alpha/\pi \sim 10^{-3}$, calculated to leading order in the large
proton mass expansion. The obtained results can be used for the
planning of experiments on the search for contributions of order
$10^{-4}$ of interactions beyond the Standard model.
\end{abstract}
\maketitle

\section{Introduction}
\label{sec:introduction}

It is well--known that the neutron $\beta^-$--decay is a good
laboratory for tests of the Standard model (SM)
\cite{Abele2008}--\cite{Ivanov2013a}. As has been pointed out in
\cite{Ivanov2013,Ivanov2013a} by example of the neutron
$\beta^-$--decay with a polarized neutron and unpolarized proton and
electron the weak magnetism and proton recoil corrections of order
$E_e/m_p$, where $E_e$ and $m_p$ are the electron energy and proton
mass, and radiative corrections of order $\alpha/\pi$, where $\alpha$
is the fine--structure constant \cite{PDG2014}, define a complete set
of corrections to the correlation coefficients of order
$10^{-3}$. These corrections provide a robust background for a search
of contributions of order $10^{-4}$, induced by interactions beyond
the Standard model (SM) \cite{Ivanov2013,Ivanov2013a}. This paper is
addressed a precision analysis of the neutron $\beta^-$--decay with
polarized neutron and electron. The aim of this paper is to give a
robust background to order $10^{-3}$ for the experimental search for
contributions of order $10^{-4}$, caused by interactions beyond the
SM. According to \cite{Ivanov2013,Ivanov2013a}, for the realization of
this aim we have to calculate in the SM the correlation coefficients
of the electron--energy spectrum and angular distribution of the
neutron $\beta^-$--decay with polarized neutron and electron by taking
into account the contributions of the weak magnetism and proton recoil
to next--to--leading order in the large proton mass expansion and
radiative corrections of order $\alpha/\pi$ to leading order in the
large proton mass expansion. These corrections make possible a
meaningful search for contributions of order $10^{-4}$, caused by
interactions beyond the SM for most of the correlation coefficients
presented here. Of course, these corrections should be meaningful if
the theoretical uncertainties of the correlation coefficients,
calculated to leading order in the large proton mass expansion and
without radiative corrections, should be small compared to
$10^{-3}$. The correlation coefficients $G(E_e)$, $N(E_e)$, $Q_e(E_e)$
and $R(E_e)$ (see Eq.(\ref{eq:3})), calculated to leading order in the
large proton mass expansion and without radiative corrections, are
equal to (see Eq.(\ref{eq:8}))
\begin{eqnarray}\label{eq:1}
\hspace{-0.3in}G_0(E_e) = - 1\;,\; N_0(E_e) = -
\frac{m_e}{E_e}\,A_0\;,\; Q_{e0}(E_e) = - A_0\;,\; R_0(E_e) = -
\alpha\,\frac{m_e}{k_e}\,A_0\;,\; A_0 = - 2\,\frac{\lambda (1 +
  \lambda)}{1 + 3 \lambda^2},
\end{eqnarray}
where $m_e$ and $k_e = \sqrt{E^2_e - m^2_e}$ are the electron mass and
its momentum, $\alpha = 1/137.036$ is the fine--structure constant
\cite{PDG2014} and $\lambda$ is the axial coupling constant
\cite{Abele2008,Nico2009}. The dependence of the correlation
coefficient $R_0(E_e)$ on the fine--structure constant is caused by
the Coulomb distortion of the electron wave function in the Coulomb
field of the proton \cite{Jackson58,Konopinski}. The most precise,
published values for the electron asymmetry from a single experiment
provide values for $A^{(\exp)}_0 = - 0.11933(34)$ \cite{Abele2008} and
$A^{(\exp)}_0 = - 0.11996(58)$ \cite{Abele2013} (see also
\cite{Abele2014}), giving $\lambda = - 1.2750(9)$ and $\lambda = -
1.2767(16)$, respectively, and averaging over the electron--energy
spectrum (see Eq.(D-59) of Ref.\cite{Ivanov2013}) we obtain the
following numerical values for the correlation coefficients in
Eq.(\ref{eq:1})
\begin{eqnarray}\label{eq:2}
\hspace{-0.3in}G_0 &=& \langle G_0(E_e)\rangle = - 1\;, \; N_0 =
\langle N_0(E_e)\rangle = 0.07825(22)\;,\nonumber\\
\hspace{-0.3in}Q_{e0} &=& \langle Q_{e0}(E_e)\rangle =
0.11933(34)\;,\; R_0 = \langle R_0(E_e)\rangle = 0.000891(3)\quad
\textrm{\cite{Abele2008}}
\;, \nonumber\\ 
\hspace{-0.3in}G_0 &=& \langle G_0(E_e)\rangle = - 1\;, \; N_0 =
\langle N_0(E_e)\rangle = 0.07866(37)\;,\nonumber\\
\hspace{-0.3in}Q_{e0} &=& \langle Q_{e0}(E_e)\rangle =
0.11996(58)\;,\; R_0 = \langle R_0(E_e)\rangle = 0.000896(5) \quad
\textrm{\cite{Abele2013}}.
\end{eqnarray}
One may see that the relative theoretical uncertainties of the
correlation coefficients in Eq.(\ref{eq:2}) are of order $(3 -
5)\times 10^{-3}$. They are practically defined by the experimental
uncertainties of the axial coupling constant. An improvement of the
theoretical uncertainties of the correlation coefficients in
Eq.(\ref{eq:2}) may only result through an improvement of the
experimental uncertainty of the axial coupling constant
$\lambda$. Hence, currently because of the theoretical uncertainties
of order $(3 - 5)\times 10^{-3}$ a precision analysis of the
correlation coefficients $G(E_e)$, $N(E_e)$, $Q_e(E_e)$ and $R(E_e)$
to order $10^{-3}$ seems to be meaningful only for the correlation
coefficient $G(E_e)$. Nevertheless, we calculate below the
corrections, caused by the weak magnetism and proton recoil of order
$1/m_p$ and radiative corrections of order $\alpha/\pi$, to all
correlation coefficients. In order to push meaningful tests of the SM
to the $10^{-4}$ level for the correlation coefficients $N(E_e)$,
$Q_e(E_e)$ and $R(E_e)$, an improvement of experimental uncertainties
for the axial coupling constant is required in addition to the
theoretical predictions we present in this work. We see this as an
important challenge for the experimental characterization of the
charged weak interaction.

The first experimental analysis of the neutron $\beta^-$--decay $n \to
p + e^- + \bar{\nu}_e$ with polarized neutron and electron has been
undertaken by Kozela {\it et al.}  \cite{Kozela2009,Kozela2012}, where
the correlation coefficients $R(E_e)$ and $N(E_e)$ of the electron
energy spectrum and angular distribution of the neutron
$\beta^-$--decay with polarized neutron and with the electron,
polarized transverse to its 3--momentum, were measured. The
correlation coefficient $R(E_e)$ describes the correlation between the
neutron polarization and the polarization and 3-momentum of the
electron $\vec{\xi}_n\cdot (\vec{k}_e\times \vec{\xi}_e)$, where
$\vec{\xi}_n$ and $\vec{\xi}_e$ are unit polarization vectors of the
neutron and electron, respectively, and $\vec{k}_e$ is the electron
3--momentum. The correlation coefficient $R$ characterizes
quantitatively a T--odd and a P--odd effect, caused by violation of time
reversal invariance and invariance under parity transformation.  The
correlation coefficient $N(E_e)$ is a quantitative characteristic of
the correlation between the neutron and electron polarization
$\vec{\xi}_n\cdot \vec{\xi}_e$. One may see that the experimental
values $R_{\exp} = 0.004 \pm 0.012\pm 0.005$ and $N_{\exp} = 0.067 \pm
0.011 \pm 0.004$, measured by Kozela {\it et al.} \cite{Kozela2012},
do not contradict the predictions of the SM, given by Eq.(\ref{eq:2}),
within the experimental uncertainties. The primary goal of the nTRV
Collaboration through these measurements of $R(E)$ and $N(E)$ was to
make a useful probe for contributions from interactions beyond the SM
(see Eqs. (7) - (11) of Ref. \cite{Kozela2012}). Such tests require a
search for deviations from the expected SM values for these
correlations. If the precision level is to be greatly improved, the
theoretical predictions for the SM values must also be refined to
include contributions from the weak magnetism, proton recoil and
radiative corrections. In this way, one can produce corrections to
order $10^{-3}$ in the correlation coefficients, and open a path to a
search for traces of interactions beyond the SM to $10^{-4}$.

The paper is organized as follows. In  section
\ref{sec:distribution} we give the electron--energy spectrum and
angular distribution of the $\beta^-$--decay of the neutron with
polarized neutron and electron. The correlation coefficients $G(E_e)$,
$N(E_e)$, $Q_e(E_e)$ and $R(E_e)$ are calculated in the SM with the
contributions of the weak magnetism and proton recoil to
next--to--leading order in the large proton mass expansion and with
radiative corrections of order $\alpha/\pi$, calculated to leading
order in the large proton mass expansion \cite{Ivanov2013}. In section
\ref{sec:wilkinson} we discuss some corrections of order $10^{-5}$ to
the correlation coefficients beyond those, which are calculated in
section \ref{sec:distribution} and which have been analysed and
discussed by Wilkinson \cite{Wilkinson1982}. In section
\ref{sec:conclusion} we discuss the obtained results and the
experimental observables. In the Appendix we calculate the
photon--electron energy spectrum, the electron--energy spectrum and
angular distributions of the radiative $\beta^-$--decay of the neutron
with polarized neutron and electron.

\section{Electron--energy spectrum and angular distribution}
\label{sec:distribution}

The electron--energy spectrum and angular distribution of the neutron
$\beta^-$--decay with polarized neutron and electron takes the form
\cite{SPT1} (see also \cite{SPT4} and \cite{Ivanov2013})
\begin{eqnarray*}
\hspace{-0.3in}\frac{d^3 \lambda_n(E_e)}{dE_e d\Omega_e} &=& (1 + 3
\lambda^2)\,\frac{G^2_F|V_{ud}|^2}{8\pi^4} \,(E_0 - E_e)^2 \sqrt{E^2_e
  - m^2_e}\, E_e\,F(E_e, Z = 1)\,\zeta(E_e)\,\Big\{1 +
A_W(E_e)\,\frac{\vec{\xi}_n\cdot \vec{k}_e}{E_e}\nonumber\\
\end{eqnarray*}
\begin{eqnarray}\label{eq:3}
\hspace{-0.3in}&& + G(E_e)\,\frac{\vec{\xi}_e \cdot \vec{k}_e}{E_e} +
N(E_e)\,\vec{\xi}_n\cdot \vec{\xi}_e +
Q_e(E_e)\,\frac{(\vec{\xi}_n\cdot \vec{k}_e)( \vec{k}_e\cdot
  \vec{\xi}_e)}{E_e (E_e + m_e)} +
R(E_e)\,\frac{\vec{\xi}_n\cdot(\vec{k}_e \times
  \vec{\xi}_e)}{E_e}\Big\},
\end{eqnarray}
where $G_F = 1.1664\times 10^{-11}\,{\rm MeV}^{-2}$ is the Fermi weak
constant, $V_{ud} = 0.97428(15)$ is the Cabibbo-Kobayashi--Maskawa
(CKM) matrix element \cite{PDG2014}, $\lambda$ is a real axial
coupling constant, $E_0 = (m^2_n - m^2_p + m^2_e)/2 m_n = 1.2927\,{\rm
  MeV}$ is the end--point energy of the electron spectrum, calculated
for $m_n = 939.5654\,{\rm MeV}$, and $m_p = 938.2720\,{\rm MeV}$ and
$m_e = 0.5110\,{\rm MeV}$ \cite{PDG2014}, $\vec{\xi}_n$ and
$\vec{\xi}_e$ are unit polarization vectors of the neutron and
electron, respectively, $F(E_e, Z = 1)$ is the relativistic Fermi
function \cite{Konopinski,Ivanov2013}
\begin{eqnarray}\label{eq:4}
\hspace{-0.3in}F(E_e, Z = 1 ) = \Big(1 +
\frac{1}{2}\gamma\Big)\,\frac{4(2 r_pm_e\beta)^{2\gamma}}{\Gamma^2(3 +
  2\gamma)}\,\frac{\displaystyle e^{\,\pi
 \alpha/\beta}}{(1 - \beta^2)^{\gamma}}\,\Big|\Gamma\Big(1 + \gamma +
 i\,\frac{\alpha }{\beta}\Big)\Big|^2,
\end{eqnarray}
where $\beta = k_e/E_e = \sqrt{E^2_e - m^2_e}/E_e$ is the electron
velocity, $\gamma = \sqrt{1 - \alpha^2} - 1$, $r_p$ is the electric
radius of the proton.  In the numerical calculations we will use $r_p
= 0.875\,{\rm fm}$ \cite{LEP}. The Fermi function Eq.(\ref{eq:4})
describes the contribution of the electron--proton final--state
Coulomb interaction. Since it is defined by the exact solution of the
Dirac equation for the electron, moving in the Coulomb field of the
proton \cite{Konopinski}, it cannot introduce additional uncertainties
to the approximate contributions, caused by the weak magnetism, proton
recoil and radiative corrections. We would like to emphasize that the
Fermi function Eq.(\ref{eq:4}) gives a contribution to the phase space
factor of the neutron of about $3.32\,\%$. The use of the approximate
expression $F(E_e, Z = 1) = 1 + \alpha\,\pi/\beta$
\cite{Wilkinson1982} diminishes the contribution of the Coulomb
electron--proton final--state interaction at the level of $8.75\times
10^{-4}$. This justifies the use of the exact Fermi function
Eq.(\ref{eq:4}) for the precision analysis of the neutron
$\beta^-$--decay.

The correlation coefficients of the electron--energy spectrum and 
angular distribution Eq.(\ref{eq:3}) we calculate with the
Hamiltonian of $V-A$ weak interactions and the weak magnetism
\cite{Ivanov2013}
\begin{eqnarray}\label{eq:5}
\hspace{-0.3in}{\cal H}_W(x) =
\frac{G_F}{\sqrt{2}}\,V_{ud}\,\Big\{[\bar{\psi}_p(x)\gamma_{\mu}(1+
  \lambda \gamma^5)\psi_n(x)] + \frac{\kappa}{2 M}
\partial^{\nu}[\bar{\psi}_p(x)\sigma_{\mu\nu}\psi_n(x)]\Big\}
        [\bar{\psi}_e(x)\gamma^{\mu}(1 - \gamma^5)\psi_{\nu_e}(x)],
\end{eqnarray}
where $\psi_p(x)$, $\psi_n(x)$, $\psi_e(x)$ and $\psi_{\nu_e}(x)$ are
the field operators of the proton, neutron, electron and
anti-neutrino, respectively, $\gamma^{\mu}$, $\sigma^{\mu\nu} =
\frac{i}{2}(\gamma^{\mu}\gamma^{\nu} - \gamma^{\nu}\gamma^{\mu})$ and
$\gamma^5$ are the Dirac matrices; $\kappa = \kappa_p - \kappa_n =
3.7058$ is the isovector anomalous magnetic moment of the nucleon,
defined by the anomalous magnetic moments of the proton $\kappa_p =
1.7928$ and the neutron $\kappa_n = - 1.9130$ and measured in nuclear
magneton \cite{PDG2014}, and $2 M = m_n + m_p$ is the average nucleon
mass.

The coefficients $\zeta(E_e)$ and $A_W(E_e)$ have been calculated in
\cite{Gudkov2006,Ivanov2013}. They read
\begin{eqnarray}\label{eq:6}
\hspace{-0.3in}\zeta(E_e) &=&\Big(1 +
\frac{\alpha}{\pi}\,g_n(E_e)\Big) + \frac{1}{M}\,\frac{1}{1 +
  3\lambda^2}\,\Big[- 2\,\Big(\lambda^2 - (\kappa +
  1)\,\lambda\Big)\,E_0 + \Big(10 \lambda^2 - 4(\kappa + 1)\,\lambda +
  2\Big)\,E_e\nonumber\\
 \hspace{-0.3in}&-& 2 \,\Big(\lambda^2 - (\kappa +
  1)\,\lambda\Big)\,\frac{m^2_e}{E_e}\Big],\nonumber\\
\hspace{-0.3in}\zeta(E_e)\,A_W(E_e) &=& \zeta(E_e)\,\Big(A(E_e) +
\frac{1}{3}\,Q_n(E_e)\Big) = A_0\,\Big(1 +
\frac{\alpha}{\pi}\,g_n(E_e) +
\frac{\alpha}{\pi}\,f_n(E_e)\Big) +  \frac{1}{M}\,\frac{1}{1 +
   3\lambda^2}\nonumber\\
\hspace{-0.3in}&&\times\,\Big[\Big(\frac{4}{3}\,\lambda^2 -
  \Big(\frac{4}{3}\kappa + \frac{2}{3}\Big)\,\lambda -
  \frac{2}{3}(\kappa + 1)\Big)\,E_0 - \Big(\frac{22}{3}\lambda^2 -
  \Big(\frac{10}{3}\kappa - \frac{4}{3}\Big)\,\lambda -
  \frac{2}{3}(\kappa + 1)\Big)\,E_e\Big],
\end{eqnarray}
where the correlation coefficients $A(E_e)$ and $Q_n(E_e)$ are given
in \cite{Gudkov2006,Ivanov2013}. The correlation coefficient
$A_W(E_e)$ without the contribution of the radiative corrections,
defined by the function $f_n(E_e)$, has been calculated by Wilkinson
\cite{Wilkinson1982}. The radiative corrections $g_n(E_e)$ and
$f_n(E_e)$ (see \cite{Ivanov2013}) are in analytical agreement with
the radiative corrections, obtained by Sirlin {\it et al.}
\cite{Sirlin1967} and Gudkov {\it et al.}  \cite{Gudkov2006},
respectively (where the function $f_n(E_e)$ was calculated for the
first time by Shann \cite{Shann1971}).

Using the results, obtained in \cite{Ivanov2013} (see Appendix A of
Ref.\cite{Ivanov2013}), for other correlation coefficients in
Eq.(\ref{eq:3}) we get the expressions
\begin{eqnarray*}
\hspace{-0.3in}\zeta(E_e)G(E_e) &=& - \Big(1 +
\frac{\alpha}{\pi}\,g_n(E_e) + \frac{\alpha}{\pi}\,f_n(E_e)\Big) +
\frac{1}{M}\,\frac{1}{1 + 3\lambda^2}\,\Big[- \Big(10 \lambda^2 -
  4(\kappa + 1)\,\lambda + 2\Big)E_e\nonumber\\
\hspace{-0.3in}&&+ \Big(2 \lambda^2 - 2 (\kappa + 1)\,
\lambda\,\Big)\,E_0\Big]\nonumber\\
\hspace{-0.3in}\zeta(E_e)N(E_e) &=& + \frac{m_e}{E_e}\,\Big\{ -
A_0\,\Big(1 + \frac{\alpha}{\pi}\,g_n(E_e) +
\frac{\alpha}{\pi}\,h^{(1)}_n(E_e)\Big) + \frac{1}{M}\,\frac{1}{1 +
  3\lambda^2}\,\Big[\Big(\frac{16}{3}\,\lambda^2 - \Big(\frac{4}{3}
  \kappa - \frac{16}{3}\Big)\,\lambda\nonumber\\
\hspace{-0.3in}&& - \frac{2}{3} (\kappa + 1)\Big) E_e -
\Big(\frac{4}{3}\,\lambda^2 - (\frac{4}{3} \kappa -
\frac{1}{3}\Big)\,\lambda - \frac{2}{3} (\kappa + 1)\Big)
E_0\Big]\Big\},\nonumber\\
\end{eqnarray*}
\begin{eqnarray}\label{eq:7}
\hspace{-0.3in}\zeta(E_e) Q_e(E_e) &=& - A_0\,\Big(1 +
\frac{\alpha}{\pi}\,g_n(E_e) + \frac{\alpha}{\pi}\,h^{(2)}_n(E_e)\Big)
+ \frac{1}{M}\,\frac{1}{1 + 3\lambda^2}\,\Big[\Big(\frac{22}{3}\,
  \lambda^2 - \Big(\frac{10}{3}\kappa - \frac{10}{3}\Big)\,\lambda -
  \frac{2}{3} (\kappa + 1)\Big) E_e\nonumber\\
\hspace{-0.3in}&& - \Big(\frac{4}{3}\,\lambda^2 - \Big(\frac{4}{3}
\kappa - \frac{1}{3}\Big)\,\lambda - \frac{2}{3} (\kappa + 1)\Big) E_0
+ \Big(2\,\lambda^2 - (2 \kappa + 1)\,\lambda\Big) m_e\Big],
\nonumber\\
\hspace{-0.3in}\zeta(E_e) R(E_e) &=& - \alpha\,\frac{m_e}{k_e}\,A_0.
\end{eqnarray}
The functions $f_n(E_e)$, $h^{(\ell)}_n(E_e)$ for $\ell = 1,2$
describe the radiative corrections of order $\alpha/\pi$. They are
calculated in the Appendix (see Eq.(\ref{eq:A.9})) and plotted in
Fig\,\ref{fig:1}. In the electron energy region $m_e \le E_e \le E_0$
they vary over the regions $2.81\times 10^{-3} \ge
(\alpha/\pi)\,f_n(E_e) \ge 6.24\times 10^{-4}$, $-6.04\times 10^{-4
}\ge (\alpha/\pi)\,h^{(1)}_n(E_e) \ge - 3.37\times 10^{-3}$ and
$5.07\times 10^{-3}\ge (\alpha/\pi)\,h^{(2)}_n(E_e) \ge 2.20\times
10^{-3}$, respectively.
\begin{figure}
\includegraphics[width=0.35\linewidth]{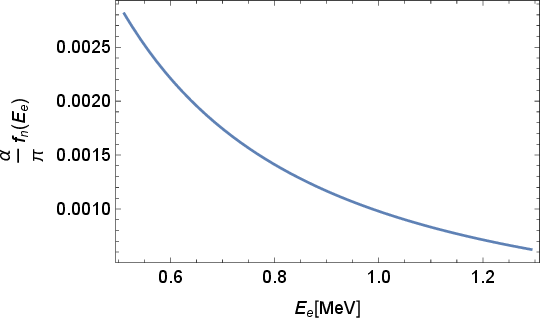}
\includegraphics[width=0.35\linewidth]{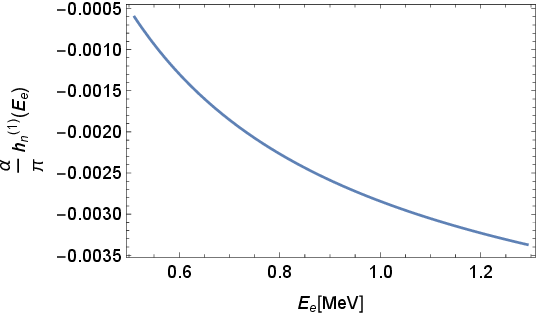}
\includegraphics[width=0.35\linewidth]{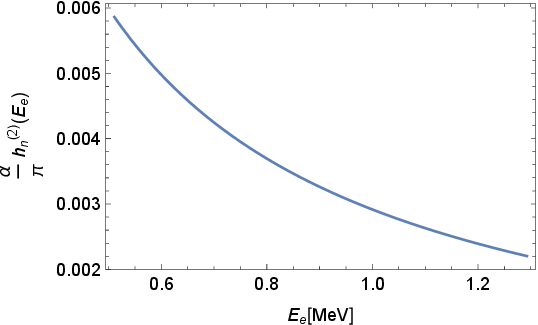}
  \caption{Radiative corrections $(\alpha/\pi)\,f_n(E_e)$,
    $(\alpha/\pi)\,h^{(1)}_n(E_e)$ and $(\alpha/\pi)\,h^{(2)}_n(E_e)$
    to the correlation coefficients $G(E_e)$, $N(E_e)$ and $Q_e(E_e)$
    of the electron--energy spectrum and angular distribution
    Eq.(\ref{eq:3}).}
\label{fig:1}
\end{figure}
The term proportional to the fine--structure constant
$\alpha$ in the correlation coefficient $\zeta(E_e)R(E_e)$ is induced
by the Coulomb distortion of the Dirac bispinor wave function of the
electron \cite{Jackson58,Konopinski}.

Keeping the contributions of the terms of order of $1/M$ inclusively
the correlation coefficients under consideration take the form
\begin{eqnarray}\label{eq:8}
\hspace{-0.3in}G(E_e) &=& - \Big(1 +
\frac{\alpha}{\pi}\,f_n(E_e)\Big)\,\Big(1 + \frac{1}{M}\,\frac{1}{1 +
  3\lambda^2}\,\Big(2 \lambda^2 - 2 (\kappa +
1)\,\lambda\Big)\,\frac{m^2_e}{E_e}\Big),\nonumber\\
\hspace{-0.3in}N(E_e) &=& + \Big(1 +
\frac{\alpha}{\pi}\,h^{(1)}_n(E_e)\Big)\,\frac{m_e}{E_e}\,\Big\{ - A_0
+ \frac{1}{M}\,\frac{1}{1 + 3
  \lambda^2}\,\Big[\Big(\frac{16}{3}\,\lambda^2 -
  \Big(\frac{4}{3}\kappa - \frac{16}{3}\Big)\,\lambda -
  \frac{2}{3}(\kappa + 1)\Big) E_e\nonumber\\
\hspace{-0.3in}&& -\Big(\frac{4}{3}\,\lambda^2 -
\Big(\frac{4}{3}\kappa - \frac{1}{3}\Big)\,\lambda -
\frac{2}{3}(\kappa + 1)\Big) E_0\Big] - \frac{1}{M}\, \frac{A_0}{1 + 3
  \lambda^2}\,\Big[- \Big(10 \lambda^2 - 4(\kappa + 1)\,\lambda +
  2\Big)\,E_e\nonumber\\
\hspace{-0.3in}&& + \Big(2 \lambda^2 - 2(\kappa +
1)\,\lambda\Big)\,\Big(E_0 +
\frac{m^2_e}{E_e}\Big)\Big]\Big\},\nonumber\\
\hspace{-0.3in}Q_e(E_e) &=& \,\Big(1 +
\frac{\alpha}{\pi}\,h^{(2)}_n(E_e)\Big)\,\Big\{ - A_0 +
\frac{1}{M}\,\frac{1}{1 + 3
  \lambda^2}\,\Big[\Big(\frac{22}{3}\,\lambda^2 - \Big(\frac{10}{3}
  \kappa - \frac{10}{3}\Big)\,\lambda - \frac{2}{3}(\kappa + 1)\Big)
  E_e\nonumber\\
\hspace{-0.3in}&&- \Big(\frac{4}{3}\,\lambda^2 - \Big(\frac{4}{3}
\kappa - \frac{1}{3}\Big)\,\lambda - \frac{2}{3}(\kappa + 1)\Big) E_0
+ \Big(2 \lambda^2 - 2 (\kappa + 1)\, \lambda\Big) m_e\Big] -
\frac{1}{M}\, \frac{A_0}{1 + 3 \lambda^2}\nonumber\\
\hspace{-0.3in}&&\,\Big[- \Big(10 \lambda^2 - 4(\kappa + 1)\,\lambda +
  2\Big)\,E_e + \Big(2 \lambda^2 - 2(\kappa +
  1)\,\lambda\Big)\,\Big(E_0 +
  \frac{m^2_e}{E_e}\Big)\Big]\Big\},\nonumber\\
\hspace{-0.3in}R(E_e) &=& - \alpha\,\frac{m_e}{k_e}\,A_0,
\end{eqnarray}
where we have neglected the terms of order $
  (\alpha/\pi)(E_e/M) < 3\times 10^{-6}$. The correlation
  coefficients Eq.(\ref{eq:8}) are defined by a complete set of
  contributions to order $10^{-3}$, caused by the weak magnetism and
  proton recoil corrections of order $1/M$ and radiative corrections
  of order $\alpha/\pi$. For example, at $\lambda = -1,2750(9)$ and
  $E_0 = 1.2927\,{\rm MeV}$ we get
\begin{eqnarray}\label{eq:9}
\hspace{-0.3in}G(E_e) &=& - \Big(1 +
\frac{\alpha}{\pi}\,f_n(E_e)\Big)\,\Big(1 + 1.41\times
10^{-3}\,\frac{m_e}{E_e}\Big),\nonumber\\
\hspace{-0.3in}N(E_e) &=& - \Big(1 +
\frac{\alpha}{\pi}\,h^{(1)}_n(E_e)\Big)\,\frac{m_e}{E_e}\,A_0\,\Big\{1
+ \Big(-\,6.06\times 10^{-3} + 1.41 \times 10^{-3}\,\frac{m_e}{E_e} -
1.85\times 10^{-5}\,\frac{E_e}{E_0}\Big)\Big\},\nonumber\\
\hspace{-0.3in}Q_e(E_e) &=& - \,\Big(1 +
\frac{\alpha}{\pi}\,h^{(2)}_n(E_e)\Big)\,A_0\Big\{1 + \Big(-
6.06\times 10^{-3} + 1.41\times 10^{-3}\,\frac{m_e}{E_e} + 2.99\times
10^{-2}\,\frac{E_e}{E_0}\Big)\Big\},
\end{eqnarray}
where the correlation coefficient $A_0$ is factorized out of the
brackets of the correlation coefficients $N(E_e)$ and $Q_e(E_e)$.  The
obtained results provide a robust theoretical background to order
$10^{-3}$ for planning experiments on the search for contributions of
order $10^{-4}$ of interactions beyond the SM. The appearance of the
term of order $10^{-5}$ is caused by an occasional cancellation of
different contributions.
 
\section{Wilkinson's analysis of higher order corrections}
\label{sec:wilkinson}

In this section we discuss the contibutions of higher order
corrections, which are not calculated in section
\ref{sec:distribution}. These corrections were calculated by Wilkinson
\cite{Wilkinson1982} and we apply them to the analysis of the
correlations coefficients $G(E_e)$, $N(E)$, $Q_e(E_e)$ and $R(E_e)$,
respectively. According to Wilkinson \cite{Wilkinson1982}, the higher
order corrections with respect to those calculated in section
\ref{sec:distribution} should be caused by i) the proton recoil in the
Coulomb electron--proton final--state interaction, ii) the finite
proton radius, iii) the proton--lepton convolution and iv) the
higher--order {\it outer} radiative corrections.

\subsection{Proton recoil corrections, caused by the
    Coulomb electron--proton final--state interaction}

\begin{figure}
\includegraphics[width=0.37\linewidth]{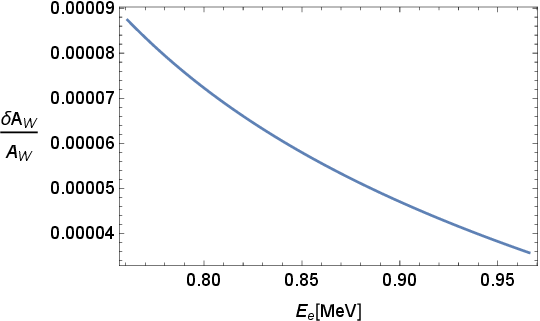}
\includegraphics[width=0.35\linewidth]{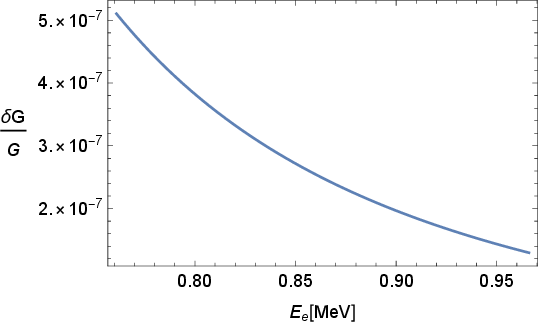}
\includegraphics[width=0.37\linewidth]{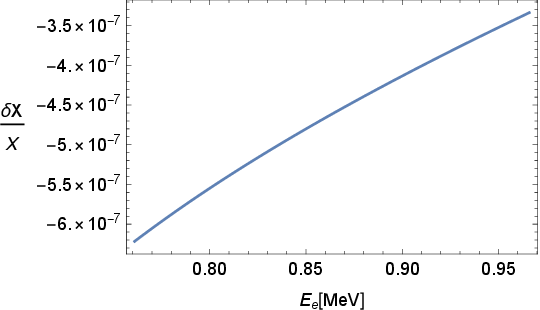}
  \caption{Relative corrections to the correlation coefficients
    $A_W(E_e)$, $G(E_e)$ and $X(E_e) = N(E_e), Q_e(E_e)$ induced by
    the proton recoil to the Fermi function, caused by the Coulomb
    electron--proton final--state interaction and calculated for the
    experimentally observable electron energy region $0.761\,{\rm MeV}
    \le E_e \le 0.966\,{\rm MeV}$ \cite{Ivanov2013}.}
\label{fig:2}
\end{figure}

As has been found by Ivanov {\it et al.}  \cite{Ivanov2013} proton
recoil, caused by the Coulomb electron--proton final--state
interaction, leads to the following change of the Fermi function
$F(E_e, Z = 1)$ (see Appendix H of Ref.\cite{Ivanov2013})
\begin{eqnarray}\label{eq:10}
\hspace{-0.3in}&&F(E_e, Z = 1) \to F(E_e, Z = 1)\,\Big(1 - \frac{\pi
  \alpha}{\beta}\,\frac{E_e}{M} - \frac{\pi
  \alpha}{\beta^3}\,\frac{E_0 - E_e}{M}\,\frac{\vec{k}_e\cdot
  \vec{k}_{\nu}}{E_e E_{\nu}}\Big),
\end{eqnarray}
where we have taken only the leading order $\alpha/M$
contributions. Then, $E_{\nu} = E_0 - E_e$ and $\vec{k}_{\nu}$ are the
energy and 3--momentum of the electron antineutrino. As has been shown
in \cite{Ivanov2013} the contribution of the proton recoil, caused by
the final--state Coulomb electron--proton interaction
Eq.(\ref{eq:10}), to the function $\zeta(E_e)$ agrees well with the
result, obtained by Wilkinson \cite{Wilkinson1982}. For the
calculation of the corrections to the correlation coefficients
$A_W(E_e)$, $G(E_e)$, $N(E_e)$ and $Q_e(E_e)$, caused by the change of
the Fermi function Eq.(\ref{eq:10}), we have to take into account the
contributions of the correlation corfficients $\zeta(E_e)$,
$a(E_e)(\vec{k}_e\cdot \vec{k}_{\nu}/E_eE_{\nu})$, $B(E_e)
(\vec{\xi}_n\cdot \vec{k}_{\nu}/E_{\nu})$, $H(E_e)(\vec{\xi}_e\cdot
\vec{k}_{\nu}/E_{\nu})$ and $K_e(E_e)(\vec{\xi}_e\cdot \vec{k}_e)(
\vec{k}_e\cdot \vec{k}_{\nu})/(E_e + m_e)E_e E_{\nu}$ and then to
integrate over the directions of the anineutrino 3--momentum
$\vec{k}_{\nu}$ \cite{Ivanov2018}. As a result we get
\begin{eqnarray}\label{eq:11}
\hspace{-0.3in}\frac{\delta A_W(E_e)}{A_W(E_e)}
&=&\frac{1}{3}\,\frac{1 - \lambda^2}{1 + 3\lambda^2}\,\frac{\pi
  \alpha}{\beta}\,\frac{E_0 - E_e}{M} - \frac{1}{3}\,\frac{1 -
  \lambda}{1 + \lambda}\,\frac{\pi \alpha}{\beta^3}\,\frac{E_0 -
  E_e}{M},\nonumber\\
\hspace{-0.3in}\frac{\delta G(E_e)}{G(E_e)} &=& - \frac{1}{3}\,\frac{1
  - \lambda^2}{1 + 3\lambda^2}\, (1 - \beta^2)\,\frac{\pi
  \alpha}{\beta^3}\,\frac{E_0 - E_e}{M},\nonumber\\
\hspace{-0.3in}\frac{\delta
  N(E_e)}{N(E_e)} &=& \frac{\delta Q_e(E_e)}{Q_e(E_e)} =
\frac{1}{3}\,\frac{1 - \lambda^2}{1 + 3\lambda^2}\,\frac{\pi
  \alpha}{\beta}\,\frac{E_0 - E_e}{M}.
\end{eqnarray}
In the experimental electron energy region $0.761\,{\rm MeV} \le E_e
\le 0.966\,{\rm MeV}$ the corrections to the correlation coefficients
are plotted in Fig.\,\ref{fig:2}. They vary in the following limits
$9.0\times 10^{-5}\ge \delta A_W(E_e)/A_W(E_e) \ge 3.8 \times
10^{-5}$, $5.1\times 10^{-7} \ge \delta G(E_e)/G(E_e) \ge 1.3 \times
10^{-7}$ and $- 6.3\times 10^{-7}\le \delta X(E_e)/X(E_e) \le - 3.5
\times 10^{-7}$ for $X(E_e) = N(E_e)$ and $Q_e(E_e)$, respectively.

\subsection{Corrections, caused by finite proton--radius}

According to Wilkinson \cite{Wilkinson1982}, the finite proton--radius
correction to the phase--space factor of the neutron $\beta^-$--decay
takes the form

\begin{eqnarray}\label{eq:12}
\hspace{-0.3in}L(E_e, Z = 1) = 1 + \frac{13}{60}\,\alpha^2 -
\alpha\,r_p\,E_e\,\Big(1 - \frac{1}{2}\,\frac{m^2_e}{E^2_e}\Big) = 1 +
1.154\times 10^{-5} - 4.183\times 10^{-5}\,\frac{E_e}{E_0} +
0.827\times 10^{-5}\,\frac{m_e}{E_e}.
\end{eqnarray}
The contribution of the function $L(E_e, Z = 1)$ can be absorbed by
the function $\zeta(E_e)$ and through the expansions Eq.(\ref{eq:8})
may provide equal corrections to the correlation coefficients
$G(E_e)$, $N(E_e)$ and $Q_e(E_e)$
\begin{eqnarray}\label{eq:13}
\hspace{-0.3in}\frac{\delta G(E_e)}{G(E_e)}= \frac{\delta
  N(E_e)}{N(E_e)} = \frac{\delta Q_e(E_e)}{Q_e(E_e)} = 1 - L(E_e, Z =
1).
\end{eqnarray}
The contribution of the finite proton--radius corrections to the
neutron lifetime is at the level of $10^{-5}$.

\subsection{Corrections, caused by lepton--nucleon convolution}

As has been pointed out by Wilkinson \cite{Wilkinson1982}, the wave
functions of the electron and electron antineutrino, calculated at the
center of the nucleon, are not constant and may undergo a distortion
in the nucleon volume that may lead to a convolution of the decay
rate. Such an effect Wilkinson has described by the function $C(E_e, Z
= 1)$. Following Wilkinson \cite{Wilkinson1982} we obtain the function
$C(E_e, Z = 1)$ in the form
\begin{eqnarray}\label{eq:14}
\hspace{-0.3in}&&C(E_e, Z = 1) = 1 + \Big[\Big(-
  \frac{9}{20}\,\alpha^2 + \frac{1}{5}\,m^2_e r^2_p -
  \frac{1}{5}\,E^2_0 r^2_p\Big) + \Big( -
  \frac{1}{5}\,\alpha\,r_p\,E_0 - \frac{2}{15}\,E^2_0
  r^2_p\Big)\,\frac{1 - \lambda^2}{1 + 3 \lambda^2}\Big]\nonumber\\
\hspace{-0.3in}&&+\Big[\Big(-
  \frac{3}{5}\,\alpha\,r_p\, E_0 + \frac{2}{5}\,E^2_0\,r^2_p\Big) +
  \Big(\frac{1}{5}\,\alpha\,r_p\, E_0 -
  \frac{2}{15}\,E^2_0\,r^2_p\Big)\, \frac{1 - \lambda^2}{1 + 3
    \lambda^2}\Big]\,\frac{E_e}{E_0} +
\frac{2}{15}\,m_e\,E_0\,r^2_p\,\frac{1 - \lambda^2}{1 + 3
  \lambda^2}\,\frac{m_e}{E_e}\nonumber\\ 
\hspace{-0.3in}&&+ \frac{2}{5}\,\Big(- 1 + \frac{1}{3}\,\frac{1 -
  \lambda^2}{1 + 3 \lambda^2}\Big)\,E^2_0 r^2_p\,\frac{E^2_e}{E^2_0} =
1 - 2.854\times 10^{-5} - 1.238\times 10^{-5}\,\frac{E_e}{E_0} -
0.018\times 10^{-5}\,\frac{m_e}{E_e}\nonumber\\ 
\hspace{-0.3in}&& - 1.361\times 10^{-5}\,\frac{E^2_e}{E^2_0}.
\end{eqnarray}
Because of the expansion Eq.(\ref{eq:8}) the corrections, caused by
the lepton--nucleon convolution, to the correlation coefficients
$G(E_e)$, $N(E_e)$ and $Q_e(E_e)$ are equal and given by
\begin{eqnarray}\label{eq:15}
\hspace{-0.3in}&&\frac{\delta G(E_e)}{G(E_e)}= \frac{\delta
  N(E_e)}{N(E_e)} = \frac{\delta Q_e(E_e)}{Q_e(E_e)} = 1 - C(E_e, Z =
1).
\end{eqnarray}
The contribution of the function $C(E_e, Z = 1)$ to the neutron
lifetime is at the level of $10^{-5}$.

\subsection{Higher--order {\it outer} radiative corrections }

The energy--independent radiative corrections of order $O(\alpha^2)$
and $O(\alpha^3)$ have been calculated by Wilkinson
\cite{Wilkinson1982}. The contribution of these corrections to the
phase--space factor Wilkinson was defined by $J(Z = 1)$. Using the
results, obtained by Wilkinson \cite{Wilkinson1982}, we get $J(Z = 1)
= 1 + 3.917\times 10^{-4}$. Of course, such corrections give equal
contributions to the correlation coefficients $\delta G(E_e)/G(E_e) =
\delta N(E_e)/N(E_e) = \delta Q_e(E_e)/Q_e(E_e) = - 3.917\times
10^{-4}$. In principle, they should be taken into account for an
experimental search of contributions of order $10^{-4}$ of
interactions beyond the SM.  The factor $J(Z = 1)$ changes the neutron
lifetime by $0.3\,{\rm s}$, which is, of course small, compared to the
current experimental accuracy of the neutron lifetime $\tau_n =
880.2(1.2)\,{\rm s}$ \cite{Arzumanov2015} (see also the world averaged
value $\tau_n = 880.2(1.0)\,{\rm s}$ \cite{PDG2016}).

\section{Conclusion}
\label{sec:conclusion}

We have calculated the correlation coefficients of the
electron--energy spectrum and angular distribution of the
$\beta^-$--decay of the neutron with polarized neutron and
electron. We have performed the calculation within the SM with $V - A$
weak interactions by taking into account the contributions of the weak
magnetism and proton recoil to next--to--leading order in the large
proton mass expansion and the radiative corrections of order
$\alpha/\pi$, calculated to leading order in the large proton mass
expansion. Such an approximation provides a theoretical background for
the analysis of contributions of order $10^{-4}$ of interactions
beyond the SM \cite{Ivanov2013,Ivanov2013a}.

The correlation coefficients $N(E_e)$ and $R(E_e)$, given by
Eq.(\ref{eq:8}), averaged over the electron energy spectrum (see
Eq.(D-59)) and calculated at $\lambda = - 1.2750(9)$, are equal to
\begin{eqnarray}\label{eq:16}
\hspace{-0.3in}\langle N(E_e)\rangle = 0.07767(22)\;,\; \langle
R(E_e)\rangle = 0.000891(3)
\end{eqnarray}
The recent experimental data $N_{\exp} = 0.067\pm 0.011\pm 0.004$ and
$R_{\exp} = 0.004 \pm 0.012 \pm 0.005$ \cite{Kozela2012} do not
contradict the predictions of the SM within the experimental
uncertainties.

Using the electron energy spectrum and angular distribution
Eq.(\ref{eq:3}) we can define the rate of the $\beta^-$--decay of
the neutron in dependence of the neutron and electron polarisations
\begin{eqnarray}\label{eq:17}
\lambda_n(\vec{\xi}_n,\vec{\xi}_e) = \lambda_n\,\Big(1 + \langle
\bar{N}(E_e)\rangle\,\vec{\xi}_n\cdot \vec{\xi}_e\Big),
\end{eqnarray}
where $\lambda_n$ is the $\beta^-$--decay rate of the neutron, defining
the lifetime of the neutron $\tau_n = 1/\lambda_n$, and equal to
\cite{Ivanov2013}
\begin{eqnarray}\label{eq:18}
  \lambda_n = (1 + 3 \lambda^2)\, \frac{G^2_F |V_{ud}|^2}{2
    \pi^3}\,f_n(E_0, Z = 1).
\end{eqnarray}
The Fermi integral $f_n(E_0, Z = 1)$ is given by \cite{Ivanov2013}
\begin{eqnarray}\label{eq:19}
\hspace{-0.3in}&& f_n(E_0, Z = 1) = \int^{E_0}_{m_e} (E_0 - E_e
)^2\,\sqrt{E^2_e - m^2_e}\,E_e \, F(E_e, Z = 1)\,\Big\{\Big(1 +
\frac{\alpha}{\pi}\,g_n(E_e)\Big) +
\frac{1}{M}\nonumber\\
\hspace{-0.3in}&& \times\,\frac{1}{1 + 3 \lambda^2}\Big[\Big(10
  \lambda^2 - 4(\kappa + 1)\,\lambda + 2\Big)\,E_e - \Big(2 \lambda^2
  - 2(\kappa + 1)\,\lambda \Big)\,\Big(E_0 +
  \frac{m^2_e}{E_e}\Big)\Big]\Big\}\, dE_e.
\end{eqnarray}
The correlation coefficient $\bar{N}(E_e)$ is defined by the
expression
\begin{eqnarray}\label{eq:20}
  \bar{N}(E_e) = N(E_e) + \frac{1}{3}\,\Big(1 -
  \frac{m_e}{E_e}\Big)\,Q_e(E_e).
\end{eqnarray}
For the experimental observation of the correlation coefficient
$\bar{N}(E_e)$ we propose to analyse the asymmetry
\begin{eqnarray}\label{eq:21}
 \hspace{-0.3in}&& P(\vec{\xi}_n,\vec{\xi}_e) =
 \frac{\lambda_n(\vec{\xi}_n,\vec{\xi}_e) - \lambda_n(- \vec{\xi}_n,
   \vec{\xi}_e)}{\lambda_n(\vec{\xi}_n,\vec{\xi}_e) + \lambda_n(-
   \vec{\xi}_n,\vec{\xi}_e)} = \langle \bar{N}(E_e)\rangle P_n P_e,
\end{eqnarray}
where $P_n$ and $P_e$ are the neutron and electron polarisations. The
asymmetry $P(\vec{\xi}_n,\vec{\xi}_e)$ can be measured for the
polarized neutron and electron with parallel and antiparallel
spins. Averaging $\bar{N}(E_e)$ over the electron energy spectrum (see
Eq.(D-59) in Appendix D of Ref. \cite{Ivanov2013}) we get $\langle
\bar{N}(E_e)\rangle = 0.0911$. Thus, the theoretical prediction for
asymmetry $P(\vec{\xi}_n,\vec{\xi}_e)$, obtained in the SM with the
weak magnetism, proton recoil and radiative corrections, is
\begin{eqnarray}\label{eq:22}
P(\vec{\xi}_n,\vec{\xi}_e) = 0.0911\,P_n P_e.
\end{eqnarray}
Our results should provide a necessary background for the measurement
of the contributions of order $10^{-4}$ to the $\beta^-$--decay of a
polarized neutron with a polarized electron, caused by interactions
beyond the SM \cite{Ivanov2013,SUSY}. 

The radiative corrections to the correlation coefficients $G(E_e)$,
$N(E_e)$ and $Q_e(E_e)$ are given by the functions $f_n(E_e)$,
$h^{(1)}_n(E_e)$ and $h^{(2)}_n(E_e)$, calculated in the Appendix. The
photon--electron and electron--energy spectra and angular
distributions of the radiative $\beta^-$--decay of the neutron with
polarized neutron and electron, obtained in the Appendix, may be used
for future experiments on the radiative $\beta^-$--decay of the
neutron \cite{Nico2006}. 

We would like to emphasize that the radiative corrections, described
by the functions $h^{(1)}_n(E_e)$ and $h^{(2)}_n(E_e)$, and the
photon--electron and electron--energy spectra and angular
distributions of the radiative $\beta^-$--decay of the neutron with
polarized neutron and electron have been never calculated in
literature before. 

Completing our discussion of these corrections, we would like to make
three comments: i) for predictions of $10^{-4}$ precision, it is
apparent that the higher order outer radiative corrections, discussed
in secion \ref{sec:wilkinson}, should be included, ii) for an
experimental search for interactions beyond the SM, a "discovery"
experiment with the required 5$\sigma$ sensitivity will require
experimental uncertainties of a few parts in $10^{-5}$, and iii) the
correction $\delta G(E_e)$, caused by the contribution of the proton
recoil to the Fermi function, can be also changed by the
contributions of the correlation coefficients $H(E_e)$ and $K_e(E_e)$
of the electron--energy and electron--antineutrino angular
distribution of the neutron $\beta^-$--decay with polrized electron
and unpolarized neutron and proton.

\section{Acknowledgements}

The results, obtained in this paper, were reported at the ``Satellite
workshop on symmetries in light and heavy flavour'', which was held on
7 - 8 November 2016 at Max Planck Institute for Astrophysics (MPA),
M\"unchen, Germany.  The work of A. N. Ivanov was supported by the
Austrian ``Fonds zur F\"orderung der Wissenschaftlichen Forschung''
(FWF) under contracts I689-N16, I862-N20, P26781-N20 and P26636-N20,
``Deutsche F\"orderungsgemeinschaft'' (DFG) AB 128/5-2 and by the \"OAW
within the New Frontiers Groups Programme, NFP 2013/09. The work of
R. H\"ollwieser was supported by the Erwin Schr\"odinger Fellowship
program of the Austrian Science Fund FWF (``Fonds zur F\"orderung der
wissenschaftlichen Forschung'') under Contract No. J3425-N27. The work
of M. Wellenzohn was supported by the MA 23 (FH-Call 16) under the
project ``Photonik - Stiftungsprofessur f\"ur Lehre''.

\section{Appendix A: Photon--electron and electron energy spectra
 and angular distributions of radiative $\beta^-$--decay of neutron
 with polarized neutron and electron}
\renewcommand{\theequation}{A-\arabic{equation}}
\setcounter{equation}{0}

Using the results, obtained in \cite{Ivanov2013}, the photon--electron
spectrum and angular distribution of the radiative $\beta^-$--decay
of the neutron with polarized neutron and electron takes the form
\begin{eqnarray}\label{eq:A.1}
\hspace{-0.3in}&&\frac{d^5\lambda_{\beta^-_c\gamma}(E_e,\omega,
  \vec{k}_e,\vec{n}\,)}{d\omega d E_e d\Omega_e d\Omega_{\gamma}} = (1
+ 3
\lambda^2)\,\frac{\alpha}{\pi}\,\frac{G^2_F|V_{ud}|^2}{(2\pi)^5}\,\sqrt{E^2_e
  - m^2_e}\,E_e\,F(E_e, Z = 1)\,(E_0 - E_e - \omega)^2\nonumber\\
\hspace{-0.3in}&&\times\,\frac{1}{\omega}\,\bigg\{\Big[\frac{\beta^2 -
    (\vec{n}\cdot \vec{\beta}\,)^2}{(1 - \vec{n}\cdot
    \vec{\beta}\,)^2}\Big(1 + \frac{\omega}{E_e}\Big) + \frac{1}{1 -
    \vec{n}\cdot \vec{\beta}}\,\frac{\omega^2}{E^2_e}\Big] +
A_0\,\vec{\xi}_n\,\cdot \Big\{\Big(\frac{\beta^2 - (\vec{n}\cdot
  \vec{\beta}\,)^2}{(1 - \vec{n}\cdot \vec{\beta}\,)^2} + \frac{1}{1 -
  \vec{n}\cdot
  \vec{\beta}}\,\frac{\omega}{E_e}\Big)\,\vec{\beta}\nonumber\\
\hspace{-0.3in}&& + \Big[- \frac{1 - \beta^2}{(1 - \vec{n}\cdot
    \vec{\beta}\,)^2}\,\frac{\omega}{E_e} + \frac{1}{1 - \vec{n}\cdot
    \vec{\beta}}\,\frac{\omega}{E_e}\,\Big(1 +
  \frac{\omega}{E_e}\Big)\Big]\,\vec{n}\,\Big\} + \Big\{-
\frac{m_e}{E_e}\Big[\frac{\beta^2 - (\vec{n}\cdot \vec{\beta}\,)^2}{(1
    - \vec{n}\cdot \vec{\beta}\,)^2}\,\zeta^0_e +
  \frac{(\vec{\beta}\cdot \vec{\zeta}_e) - (\vec{n}\cdot
    \vec{\beta}\,)(\vec{n}\cdot \vec{\zeta}_e)}{(1 - \vec{n}\cdot
    \vec{\beta}\,)^2}\,\frac{\omega}{E_e}\nonumber\\
\hspace{-0.3in}&& + \frac{\zeta^0_e - \vec{n}\cdot \vec{\zeta}_e}{(1 -
  \vec{n}\cdot \vec{\beta}\,)^2}\,\frac{\omega^2}{E^2_e}\Big] -
\frac{m_e}{E_e}\,A_0\Big[\frac{\beta^2 - (\vec{n}\cdot
    \vec{\beta}\,)^2}{(1 - \vec{n}\cdot
    \vec{\beta}\,)^2}\,(\vec{\xi}_n\cdot \vec{\zeta}_e) +
  \frac{(\vec{\beta}\cdot \vec{\zeta}_e) - (\vec{n}\cdot
    \vec{\beta}\,)(\vec{n}\cdot \vec{\zeta}_e)}{(1 - \vec{n}\cdot
    \vec{\beta}\,)^2}\,(\vec{\xi}_n\cdot
  \vec{n}\,)\,\frac{\omega}{E_e} + \frac{\zeta^0_e - \vec{n}\cdot
    \vec{\zeta}_e}{(1 - \vec{n}\cdot \vec{\beta}\,)^2}\nonumber\\
\hspace{-0.3in}&&\times\,(\vec{\xi}_n\cdot
\vec{n}\,)\,\frac{\omega^2}{E^2_e} + \frac{\omega}{E_e}\,
\frac{\big(\vec{\xi}_n \cdot \vec{\beta} - (\vec{\xi}_n \cdot
  \vec{n}\,)(\vec{n}\cdot \vec{\beta}\,)\big)(\zeta^0_e - \vec{n}\cdot
  \vec{\zeta}_e)}{(1 - \vec{n}\cdot
  \vec{\beta}\,)^2}\Big]\Big\}\bigg\},
\end{eqnarray}
where $\beta = \sqrt{E^2_e - m^2_e}/E_e$ is the electron velocity and
$\omega$ is the photon energy, the vector $\vec{n}$ is directed along
the photon 3--momentum, $d\Omega_e$ and $d\Omega_{\gamma}$ are the
elements of the solid angles of the electron and the photon,
respectively.  The 4--vector of the electron polarization
$\zeta^{\mu}_e = (\zeta^0_e, \vec{\zeta}_e)$ is defined by
\begin{eqnarray}\label{eq:A.2}
\zeta^{\mu}_e = (\zeta^0_e, \vec{\zeta}_e) =\Big(\frac{\vec{k}_e\cdot
  \vec{\xi}_e}{m_e}, \vec{\xi}_e + \frac{\vec{k}_e(\vec{k}_e\cdot
  \vec{\xi}_e)}{m_e(E_e + m_e)}\Big).
\end{eqnarray}
It obeys the constraints $\zeta^2_e = -1$ and $k_e\cdot \zeta_e = 0$.  For
the derivation of the electron energy spectrum and angular
distribution it is convenient to rewrite Eq.(\ref{eq:A.1}) as
follows
\begin{eqnarray}\label{eq:A.3}
\hspace{-0.21in}&&\frac{d^5\lambda_{\beta^-_c\gamma}(E_e,\omega,
  \vec{k}_e,\vec{n}\,)}{d\omega d E_e d\Omega_e d\Omega_{\gamma}} = (1
+ 3 \lambda^2) \frac{\alpha}{\pi} \frac{G^2_F|V_{ud}|^2}{(2\pi)^5}
\sqrt{E^2_e - m^2_e}\,E_e\,F(E_e, Z = 1)\,(E_0 - E_e - \omega)^2
\frac{1}{\omega}\bigg\{\Big[\frac{\beta^2 - (\vec{n}\cdot
    \vec{\beta}\,)^2}{(1 - \vec{n}\cdot \vec{\beta}\,)^2}\Big(1 +
  \frac{\omega}{E_e}\Big)\nonumber\\
\hspace{-0.3in}&& + \frac{1}{1 - \vec{n}\cdot
  \vec{\beta}}\,\frac{\omega^2}{E^2_e}\Big] + A_0\,
\Big\{\Big[\frac{\beta^2 - (\vec{n}\cdot \vec{\beta}\,)^2}{(1 -
    \vec{n}\cdot \vec{\beta}\,)^2} + \frac{1}{1 - \vec{n}\cdot
    \vec{\beta}}\,\frac{\omega}{E_e}\Big]\,\vec{\xi}_n\cdot\vec{\beta}
+ \Big[- \frac{1 - \beta^2}{(1 - \vec{n}\cdot
    \vec{\beta}\,)^2}\,\frac{\omega}{E_e} + \frac{1}{1 - \vec{n}\cdot
    \vec{\beta}}\,\frac{\omega}{E_e}\,\Big(1 +
  \frac{\omega}{E_e}\Big)\Big]\,\vec{\xi}_n \cdot\vec{n}\,\Big\}
\nonumber\\
\hspace{-0.3in}&& - \frac{m_e}{E_e}\Big\{\frac{\beta^2 - (\vec{n}\cdot
  \vec{\beta}\,)^2}{(1 - \vec{n}\cdot \vec{\beta}\,)^2} \zeta^0_e +
\frac{1}{(1 - \vec{n}\cdot \vec{\beta}\,)^2} \frac{\omega}{E_e}
\zeta^0_e + \Big[\frac{\vec{n}\cdot \vec{\zeta}_e}{1 - \vec{n}\cdot
    \vec{\beta}} - \frac{\vec{n}\cdot \vec{\zeta}_e}{(1 - \vec{n}\cdot
    \vec{\beta}\,)^2}\Big] \frac{\omega}{E_e} + \frac{1}{(1 -
  \vec{n}\cdot \vec{\beta}\,)^2}\frac{\omega^2}{E^2_e} \zeta^0_e -
\frac{\vec{n}\cdot \vec{\zeta}_e}{(1 - \vec{n}\cdot
  \vec{\beta}\,)^2}\frac{\omega^2}{E^2_e}\Big\}\nonumber\\
\hspace{-0.3in}&& - \frac{m_e}{E_e}\,A_0\Big\{\frac{\beta^2 -
  (\vec{n}\cdot \vec{\beta}\,)^2}{(1 - \vec{n}\cdot
  \vec{\beta}\,)^2}\,\vec{\xi}_n\cdot \vec{\zeta}_e +
\frac{\vec{n}\cdot \vec{\xi}_n}{1 - \vec{n}\cdot
  \vec{\beta}}\,\frac{\omega}{E_e}\,\zeta^0_e + \frac{\vec{n}\cdot
  \vec{\xi}_n}{(1 - \vec{n}\cdot
  \vec{\beta}\,)^2}\,\frac{\omega^2}{E^2_e}\,\zeta^0_e +
\frac{\vec{\xi}_n\cdot \vec{\beta}}{(1 - \vec{n}\cdot
  \vec{\beta}\,)^2}\, \frac{\omega}{E_e}\,\zeta^0_e -
\frac{(\vec{\beta}\cdot \vec{\xi}_n)(\vec{n}\cdot \vec{\zeta}_e)}{(1 -
  \vec{n}\cdot \vec{\beta}\,)^2}\, \frac{\omega}{E_e}\nonumber\\
\hspace{-0.3in}&& - \frac{(\vec{n}\cdot \vec{\xi}_n)(\vec{n}\cdot
  \vec{\zeta}_e)}{(1 - \vec{n}\cdot
  \vec{\beta}\,)^2}\,\frac{\omega^2}{E^2_e} \Big\}\bigg\}.
\end{eqnarray}
The integration over the directions of $\vec{n}$ we carry out with the
following auxiliary integrals
\begin{eqnarray}\label{eq:A.4}
\int \frac{\beta^2 - (\vec{n}\cdot \vec{\beta}\,)^2}{(1 - \vec{n}\cdot
  \vec{\beta}\,)^2}\,\frac{d\Omega_{\gamma}}{4\pi} &=&
\frac{1}{\beta}\,{\ell n}\Big(\frac{1+ \beta}{1- \beta}\Big) -
2\;,\;\int \frac{1}{(1 - \vec{n}\cdot
  \vec{\beta}\,)^2}\,\frac{d\Omega_{\gamma}}{4\pi} = \frac{1}{1 -
  \beta^2}\;,\; \int \frac{1}{1 - \vec{n}\cdot
  \vec{\beta}}\,\frac{d\Omega_{\gamma}}{4\pi} =
\frac{1}{2\beta}\,{\ell n}\Big(\frac{1+ \beta}{1-
  \beta}\Big),\nonumber\\ \int \frac{\vec{a}\cdot \vec{n}}{(1 -
  \vec{n}\cdot \vec{\beta}\,)^2}\,\frac{d\Omega_{\gamma}}{4\pi} &=&
-\,\frac{1}{2\beta^2}\,\Big[\frac{1}{\beta}{\ell n}\Big(\frac{1+
    \beta}{1- \beta}\Big) - \frac{2}{1 - \beta^2}\Big](\vec{a}\cdot
\vec{\beta}\,)\;,\; \int \frac{\vec{a}\cdot \vec{n}}{1 - \vec{n}\cdot
  \vec{\beta}}\,\frac{d\Omega_{\gamma}}{4\pi} =
\frac{1}{2\beta^2}\,\Big[\frac{1}{\beta}{\ell n}\Big(\frac{1+
    \beta}{1- \beta}\Big) - 2\Big](\vec{a}\cdot
\vec{\beta}\,),\nonumber\\ \int \frac{(\vec{a}\cdot
  \vec{n}\,)(\vec{b}\cdot \vec{n}\,)}{(1 - \vec{n}\cdot
  \vec{\beta}\,)^2}\,\frac{d\Omega_{\gamma}}{4\pi} &=&
\frac{1}{2}\,\frac{1}{\beta^2}\Big[\frac{1}{\beta}{\ell n}\Big(\frac{1
    + \beta}{1- \beta}\Big) - 2\Big](\vec{a}\cdot \vec{b}\,) -
\frac{1}{2}\,\frac{1}{\beta^4}\Big[\frac{3}{\beta}{\ell
    n}\Big(\frac{1+ \beta}{1- \beta}\Big) - 4 - \frac{2}{1 -
    \beta^2}\Big](\vec{a}\cdot \vec{\beta}\,)(\vec{b}\cdot
\vec{\beta}\,),\nonumber\\ \int \frac{(\vec{a}\cdot
  \vec{n}\,)(\vec{b}\cdot \vec{n}\,)}{1 - \vec{n}\cdot
  \vec{\beta}}\,\frac{d\Omega_{\gamma}}{4\pi} &=& -
\frac{1}{4}\,\frac{1}{\beta^2}\,\Big[\frac{1 - \beta^2}{\beta}{\ell
    n}\Big(\frac{1 + \beta}{1- \beta}\Big) - 2\Big](\vec{a}\cdot
\vec{b}\,) + \frac{1}{4}\,\frac{1}{\beta^4}\Big[\frac{3 -
    \beta^2}{\beta}{\ell n}\Big(\frac{1 + \beta}{1- \beta}\Big) - 6
  \Big](\vec{a}\cdot \vec{\beta}\,)(\vec{b}\cdot \vec{\beta}\,).
\end{eqnarray}
As a result the photon--electron energy spectrum and angular
distribution takes the form
\begin{eqnarray}\label{eq:A.5}
\hspace{-0.3in}&&\frac{d^4\lambda_{\beta^-_c\gamma}(E_e,
  \omega)}{d\omega d E_e d\Omega_e} = (1 + 3
\lambda^2)\,\frac{\alpha}{\pi}\,\frac{G^2_F|V_{ud}|^2}{8\pi^4}\,\sqrt{E^2_e
  - m^2_e}\,E_e\,F(E_e, Z = 1)\,(E_0 - E_e -
\omega)^2\,\frac{1}{\omega}\,\Bigg\{\Big\{\Big(1 + \frac{\omega}{E_e}
+ \frac{1}{2}\,\frac{\omega^2}{E^2_e}\Big)\nonumber\\
\hspace{-0.3in}&&\times\,\Big[\frac{1}{\beta}\,{\ell n}\Big(\frac{1 +
    \beta}{1 - \beta}\Big) - 2\Big] + \frac{\omega^2}{E^2_e}\Big\} +
\frac{\vec{k}_e\cdot \vec{\xi}_n}{E_e}\,A_0\Big(1 +
\frac{1}{\beta^2}\,\frac{\omega}{E_e} + \frac{1}{2
  \beta^2}\,\frac{\omega^2}{E^2_e}\Big)\, \Big[\frac{1}{\beta}\,{\ell
    n}\Big(\frac{1 + \beta}{1 - \beta}\Big) - 2\Big] -
\frac{\vec{k}_e\cdot \vec{\xi}_e}{E_e}\,\Big(1 +
\frac{1}{\beta^2}\,\frac{\omega}{E_e} \nonumber\\
\hspace{-0.3in}&& + \frac{1}{2
  \beta^2}\,\frac{\omega^2}{E^2_e}\Big)\Big[\frac{1}{\beta}\,{\ell
    n}\Big(\frac{1 + \beta}{1 - \beta}\Big) - 2\Big] -
(\vec{\xi}_n\cdot \vec{\zeta}_e)\, \frac{m_e}{E_e}\,A_0\, \Big(1 -
\frac{1}{2\beta^2}\,\frac{\omega^2}{E^2_e}\Big)\,\Big[\frac{1}{\beta}\,
  {\ell n}\Big(\frac{1 + \beta}{1 - \beta}\Big) - 2\Big] \nonumber\\
\hspace{-0.3in}&& - \frac{m_e}{E_e}\,(\vec{\xi}_n\cdot
\vec{\beta}\,)\zeta^0_e\,A_0\,\Big\{\frac{1}{\beta^2}\,
\frac{\omega}{E_e}\, \Big[\frac{1}{\beta}\,{\ell n}\Big(\frac{1 +
    \beta}{1 - \beta}\Big) - 2\Big] + \frac{1}{2 \beta^2}\,
\frac{\omega^2}{E^2_e} \, \Big(\frac{3 -
  \beta^2}{\beta^2}\Big[\frac{1}{\beta}\,{\ell n}\Big(\frac{1 +
    \beta}{1 - \beta}\Big) - 2\Big] - 2\Big)\Big\}\bigg\}.
\end{eqnarray}
In terms of the irreducible scalar products the photon--electron
energy spectrum and angular distribution of the radiative
$\beta^-$--decay of the neutron reads
\begin{eqnarray*}%\label{eq:A.6}
\hspace{-0.3in}&&\frac{d^4\lambda_{\beta^-_c\gamma}(E_e,
  \omega)}{d\omega d E_e d\Omega_e} = (1 + 3
\lambda^2)\,\frac{\alpha}{\pi}\,\frac{G^2_F|V_{ud}|^2}{8\pi^4}\,\sqrt{E^2_e
  - m^2_e}\,E_e\,F(E_e, Z = 1)\,(E_0 - E_e -
\omega)^2\,\frac{1}{\omega}\,\bigg\{\Big\{\Big(1 + \frac{\omega}{E_e}
+ \frac{1}{2}\,\frac{\omega^2}{E^2_e}\Big)\nonumber\\
\hspace{-0.3in}&&\times\,\Big[\frac{1}{\beta}\,{\ell n}\Big(\frac{1 +
    \beta}{1 - \beta}\Big) - 2\Big] + \frac{\omega^2}{E^2_e}\Big\} +
\frac{\vec{k}_e\cdot \vec{\xi}_n}{E_e}\,A_0\Big(1 +
\frac{1}{\beta^2}\,\frac{\omega}{E_e} + \frac{1}{2
  \beta^2}\,\frac{\omega^2}{E^2_e}\Big)\, \Big[\frac{1}{\beta}\,{\ell
    n}\Big(\frac{1 + \beta}{1 - \beta}\Big) - 2\Big] -
\frac{\vec{k}_e\cdot \vec{\xi}_e}{E_e}\Big(1 +
\frac{1}{\beta^2}\,\frac{\omega}{E_e}\nonumber\\
\hspace{-0.3in}&& + \frac{1}{2
  \beta^2}\,\frac{\omega^2}{E^2_e}\Big)\Big[\frac{1}{\beta}\,{\ell
    n}\Big(\frac{1 + \beta}{1 - \beta}\Big) - 2\Big] -
(\vec{\xi}_n\cdot \vec{\xi}_e)\, \frac{m_e}{E_e}\,A_0 \Big(1 -
\frac{1}{2\beta^2}\,\frac{\omega^2}{E^2_e}\Big)\Big[\frac{1}{\beta}\,{\ell
    n}\Big(\frac{1 + \beta}{1 - \beta}\Big) - 2\Big] -
\frac{(\vec{\xi}_n\cdot \vec{k}_e)(\vec{k}_e \cdot \vec{\xi}_e)}{ (E_e
  + m_e)E_e}\,A_0 \nonumber\\
\end{eqnarray*}
\begin{eqnarray}\label{eq:A.6}
\hspace{-0.3in}&& \times \,\Big\{\Big(1 -
\frac{1}{2\beta^2}\,\frac{\omega^2}{E^2_e}\Big)\Big[\frac{1}{\beta}{\ell
    n}\Big(\frac{1 + \beta}{1 - \beta}\Big) - 2\Big] + (1 + \sqrt{1 -
  \beta^2})\,\Big[\frac{1}{\beta^2}\, \frac{\omega}{E_e}\,
\Big[\frac{1}{\beta}\,{\ell n}\Big(\frac{1 + \beta}{1 - \beta}\Big) -
  2\Big] + \frac{1}{2 \beta^2}\, \frac{\omega^2}{E^2_e}\nonumber\\
\hspace{-0.3in}&& \times \Big(\frac{3 -
  \beta^2}{\beta^2}\Big[\frac{1}{\beta}\,{\ell n}\Big(\frac{1 +
    \beta}{1 - \beta}\Big) - 2\Big] - 2\Big)\Big]\Big\}\bigg\}.
\end{eqnarray}
Integrating over the photon energy over the region $\omega_{\rm min}
\le \omega \le E_0 - E_e$ we obtain the electron energy spectrum and
angular distribution
\begin{eqnarray}\label{eq:A.7}
\hspace{-0.3in}&&\frac{d^3\lambda_{\beta^-_c\gamma}(E_e)}{d E_e
  d\Omega_e} = (1 + 3
\lambda^2)\,\frac{\alpha}{\pi}\,\frac{G^2_F|V_{ud}|^2}{8\pi^4}\,\sqrt{E^2_e
  - m^2_e}\,E_e\,F(E_e, Z = 1)\,(E_0 - E_e)^2\nonumber\\
\hspace{-0.3in}&&\times\,\Big\{g^{(1)}_{\beta^-_c\gamma}(E_e,\omega_{\rm
  min}) + \frac{\vec{k}_e\cdot \vec{\xi}_n}{E_e}\,A_0\,
g^{(2)}_{\beta^-_c\gamma}(E_e,\omega_{\rm min}) - \frac{\vec{k}_e\cdot
  \vec{\xi}_e}{E_e}\,g^{(2)}_{\beta^-_c\gamma}(E_e,\omega_{\rm
  min})\nonumber\\
\hspace{-0.3in}&& - \,\vec{\xi}_n\cdot \vec{\xi}_e\,
\frac{m_e}{E_e}\,A_0\,g^{(3)}_{\beta^-_c\gamma}(E_e,\omega_{\rm min})-
\frac{(\vec{\xi}_n\cdot \vec{k}_e)(\vec{k}_e \cdot \vec{\xi}_e)}{E_e
  (E_e + m_e)}\,A_0\,g^{(4)}_{\beta^-_c\gamma}(E_e,\omega_{\rm
  min})\Big\},
\end{eqnarray}
where the functions $g^{(i)}_{\beta^-_c\gamma}(E_e,\omega_{\rm min})$
for $i = 1,2,3,4$ are defined by the integrals
\begin{eqnarray}\label{eq:A.8}
\hspace{-0.3in}g^{(1)}_{\beta^-_c\gamma}(E_e,\omega_{\rm min}) &=&
\int^{E_0 - E_e}_{\omega_{\rm min}}\frac{d\omega}{\omega}\frac{(E_0 -
  E_e - \omega)^2}{(E_0 - E_e)^2}\,\Big\{\Big(1 + \frac{\omega}{E_e} +
\frac{1}{2}\frac{\omega^2}{E^2_e}\Big)\,\Big[\frac{1}{\beta}\,{\ell
    n}\Big(\frac{1 + \beta}{1 - \beta}\Big) - 2\Big] +
\frac{\omega^2}{E^2_e}\Big\},\nonumber\\
\hspace{-0.3in}g^{(2)}_{\beta^-_c\gamma}(E_e,\omega_{\rm min}) &=&
\int^{E_0 - E_e}_{\omega_{\rm min}}\frac{d\omega}{\omega}\,\frac{(E_0
- E_e - \omega)^2}{(E_0 - E_e)^2}\,\Big(1 +
\frac{1}{\beta^2}\frac{\omega}{E_e} +
\frac{1}{2\beta^2}\frac{\omega^2}{E^2_e}\Big)\,\Big[\frac{1}{\beta}\,{\ell
n}\Big(\frac{1 + \beta}{1 - \beta}\Big) - 2\Big],\nonumber\\
\hspace{-0.3in}g^{(3)}_{\beta^-_c\gamma}(E_e,\omega_{\rm min}) &=&
\int^{E_0 - E_e}_{\omega_{\rm min}}\frac{d\omega}{\omega}\,\frac{(E_0
  - E_e - \omega)^2}{(E_0 - E_e)^2}\, \Big(1 -
\frac{1}{2\beta^2}\,\frac{\omega^2}{E^2_e}\Big)\Big[\frac{1}{\beta}\,{\ell
    n}\Big(\frac{1 + \beta}{1 - \beta}\Big) - 2\Big],\nonumber\\
\hspace{-0.3in}g^{(4)}_{\beta^-_c\gamma}(E_e,\omega_{\rm min}) &=&
\int^{E_0 - E_e}_{\omega_{\rm min}}\frac{d\omega}{\omega}\,\frac{(E_0
  - E_e - \omega)^2}{(E_0 - E_e)^2}\,\Big\{\Big(1 -
\frac{1}{2\beta^2}\,\frac{\omega^2}{E^2_e}\Big)\Big[\frac{1}{\beta}{\ell
    n}\Big(\frac{1 + \beta}{1 - \beta}\Big) - 2\Big]\nonumber\\
\hspace{-0.3in}&& + (1 + \sqrt{1 -
  \beta^2})\,\Big\{\frac{1}{\beta^2}\, \frac{\omega}{E_e}\,
\Big[\frac{1}{\beta}\,{\ell n}\Big(\frac{1 + \beta}{1 - \beta}\Big) -
  2\Big] + \frac{1}{2 \beta^2}\, \frac{\omega^2}{E^2_e} \Big(\frac{3 -
  \beta^2}{\beta^2}\Big[\frac{1}{\beta}\,{\ell n}\Big(\frac{1 +
    \beta}{1 - \beta}\Big) - 2\Big] - 2\Big)\Big\}.
\end{eqnarray}
In terms of the functions $g^{(i)}_{\beta^-_c\gamma}(E_e,\omega_{\rm
  min})$, depending on the infrared cut--off $\omega_{\rm min}$, for
$i = 1,2,3,4$ we determine the functions $f_n(E_e)$ and
$h^{(\ell)}_n(E_e)$ for $\ell = 1,2$, which do not depend on the
infrared cut--off $\omega_{\rm min}$. They are
\begin{eqnarray}\label{eq:A.9}
\hspace{-0.3in}f_n(E_e) &=&\lim_{\omega_{\rm min} \to
  0}[g^{(2)}_{\beta^-_c\gamma}(E_e,\omega_{\rm min}) -
  g^{(1)}_{\beta^-_c\gamma}(E_e,\omega_{\rm min})] +
g_F(E_e)\,\frac{m_e}{E_e} = \frac{1}{3}\,\frac{1 - \beta^2}{\beta^2}
\frac{E_0 - E_e}{E_e} \Big(1 + \frac{1}{8}\,\frac{E_0 -
  E_e}{E_e}\Big)\nonumber\\
\hspace{-0.3in}&&\times\,\Big[\frac{1}{\beta}\,{\ell n}\Big(\frac{1 +
    \beta}{1 - \beta}\Big) - 2\Big] - \frac{1}{12}\,\frac{(E_0 -
  E_e)^2}{E^2_e} + \frac{1 - \beta^2}{2 \beta}\,{\ell n}\Big(\frac{1 +
  \beta}{1 - \beta}\Big),\nonumber\\
\hspace{-0.3in}h^{(1)}_n(E_e) &=& \lim_{\omega_{\rm min} \to
  0}[g^{(3)}_{\beta^-_c\gamma}(E_e,\omega_{\rm min}) -
  g^{(1)}_{\beta^-_c\gamma}(E_e,\omega_{\rm min})] +
g_F(E_e)\,\frac{m_e}{E_e} - g_F(E_e)\,\frac{E_e}{m_e} = -
\frac{1}{3}\,\frac{E_0 - E_e}{E_e}\Big\{\Big(1 + \frac{1 +
  \beta^2}{8\beta^2}\nonumber\\
\hspace{-0.3in}&& \times \,\frac{E_0 - E_e}{E_e}\Big)
\Big[\frac{1}{\beta}\,{\ell n}\Big(\frac{1 + \beta}{1 - \beta}\Big) -
  2\Big] + \frac{1}{4}\, \frac{E_0 - E_e}{E_e}\Big\} -
\frac{\beta}{2}\,{\ell n}\Big(\frac{1 + \beta}{1 - \beta}\Big)
,\nonumber\\
\hspace{-0.3in}h^{(2)}_n(E_e) &=&  \lim_{\omega_{\rm min} \to
0}[g^{(4)}_{\beta^-_c\gamma}(E_e,\omega_{\rm
min}) - g^{(1)}_{\beta^-_c\gamma}(E_e,\omega_{\rm min})] + g_F(E_e)\,
\frac{m_e}{E_e} + g_F(E_e) = -
\frac{1}{3}\,\frac{E_0 - E_e}{E_e}\Big\{\Big(1 + \frac{1 +
  \beta^2}{8\beta^2}\nonumber\\
\hspace{-0.3in}&& \times \,\frac{E_0 - E_e}{E_e}\Big)
\Big[\frac{1}{\beta}\,{\ell n}\Big(\frac{1 + \beta}{1 - \beta}\Big) -
  2\Big] + \frac{1}{4}\, \frac{E_0 - E_e}{E_e}\Big\}+ \big(1 + \sqrt{1
  - \beta^2}\big) \, \Big\{\frac{1}{3}\, \frac{E_0 - E_e}{\beta^2
  E_e}\Big[\frac{1}{\beta}\,{\ell n}\Big(\frac{1 + \beta}{1 -
    \beta}\Big) - 2\Big] \nonumber\\
\hspace{-0.3in}&+& \frac{1}{24}\, \frac{(E_0 - E_e)^2}{\beta^2 E^2_e}
\Big(\frac{3 - \beta^2}{\beta^2}\Big[\frac{1}{\beta}\,{\ell
    n}\Big(\frac{1 + \beta}{1 - \beta}\Big) - 2\Big] - 2\Big) +
\frac{\sqrt{1 - \beta^2}}{2 \beta}\, {\ell n}\Big(\frac{1 + \beta}{1 -
  \beta}\Big)\Big\}.
\end{eqnarray}
For the calculation of the functions $f_n(E_e)$ and
$h^{(\ell)}_n(E_e)$ for $\ell = 1,2$, defining the radiative
corrections to the correlation coefficients $A_W(E_e)$, $G(E_e)$,
$N(E_e)$ and $Q_e(E_e)$, respectively, we have to take into account
the contribution of the virtual photon exchanges, inducing the scalar
and tensor weak nucleon--lepton coupling constant \cite{Ivanov2013}
(see Appendix B).

Finally we would like to note that for the calculation of the
radiative corrections, defined by the functions $f_n(E_e)$ and
$h^{(\ell)}_n(E_e)$ for $\ell = 1,2$, the final result does not depend
on the regularization procedure. Indeed, one may use the infrared
cut--off $\omega_{\rm min}$, which may be identified with the
experimental threshold energy of photons, and the finite--photon mass
(FPM) regularization  \cite{Sirlin1967} (see also
\cite{Gudkov2006,Ivanov2013}). In turn the function $g_n(E_e)$ has to
be calculated with the FPM regularization in order to satisfy gauge
invariance and the Kinoshita--Lee--Nauenberg theorem \cite{Sirlin1967}
(see also \cite{Ivanov2013}).


\begin{thebibliography}{9}
\bibitem{Abele2008}
H. Abele, Progr. Part. Nucl. Phys. {\bf 60}, 1
  (2008).
\bibitem{Nico2009} J. S. Nico, J. Phys. G: Nucl. Part. Phys. {\bf 36},
  104001 (2009).
\bibitem{Gudkov2006} 
V. Gudkov, G. I. Greene, and J. R. Calarco,
Phys. Rev. C {\bf 73}, 035501 (2006);
V. Gudkov, Phys. Rev. C {\bf 77}, 045502 (2008).
\bibitem{Ivanov2009} 
M. Faber, A. N. Ivanov, J. Marton, M. Pitschmann, A. P. Serebrov, 
N. I. Troitskaya,and  M. Wellenzohn,
Phys. Rev. C {\bf 80}, 035503 (2009).
\bibitem{UMW1}
T. Bhattachatya {\it et al.},
Phys. Rev. D {\bf 85}, 054512 (2012).
\bibitem{UMW2}
V. Cirigliano, J. Jenkins, and M. Gonz$\acute{\rm a}$les-Alonso,
Nucl. Phys. B {\bf 830}, 95 (2010);
V. Cirigliano, M. Gonz$\acute{\rm a}$les-Alonso, and M. L. Graesser,
 arXiv: 1210.4553 [hep-ph].
\bibitem{Ivanov2013} 
A. N. Ivanov, M. Pitschmann, and N. I. Troitskaya, 
Phys. Rev. D {\bf 88}, 073002 (2013); arXiv:1212.0332 [hep--ph].
\bibitem{Ivanov2013a}
A. N. Ivanov, R. H\"ollwieser, N. I. Troitskaya, and M. Wellenzohn,
Phys. Rev. D {\bf 88}, 065026 (2013).
\bibitem{PDG2014} 
K. A. Olive {\it et al.} (Particle Data Group),
  Chin. Phys. A {\b 38}, 090001 (2014).
\bibitem{Jackson58}
J. D. Jackson, S. B. Treiman, and H. W. Wyld Jr.,
Nucl. Phys. {\bf 4}, 206 (1957);
Z. Phys. {\bf 150}, 640 (1958).
\bibitem{Konopinski} E. J. Konopinski, in {\it The theory of beta
  radioactivity}, Oxford, At the Clarendon Press, 1966.
\bibitem{Abele2013} 
D. Mund, B. Maerkisch, M. Deissenroth, J. Krempel,
  M. Schumann, H. Abele, A. Petoukhov, and T. Soldner,
  Phys. Rev. Lett. {\bf 110}, 172502 (2013).
\bibitem{Abele2014}
B. Maerkisch and H. Abele, arXiv:1410.4220 [hep-ph], Presented at 
``8th International Workshop on the CKM Unitarity Triangle (CKM 2014)'', 
Vienna, Austria, September 8 - 12, 2014.
\bibitem{Kozela2009}
A. Kozela {\it et al.}, Phys. Rev. Lett. {\bf 102}, 172301 (2009).
\bibitem{Kozela2012}
A. Kozela {\it et al.},
Phys. Rev. C {\bf 85}, 045501 (2012).
\bibitem{Wilkinson1982} 
D. H. Wilkinson, Nucl. Phys. A {\bf 377}, 474
  (1982).
\bibitem{SPT1}
J. D. Jackson, S. B. Treiman, and H. W. Wyld Jr.,
Phys. Rev. {\bf 106}, 517 (1957).
\bibitem{SPT4}
N. Severijns, M. Beck, and O. Naviliat-Cuncic
Rev. Mod. Phys. {\bf 78}, 991 (2006).
\bibitem{LEP}
P. J. Mohr, B. N. Taylor,
Rev. Mod. Phys. {\bf 77}, 1 (2005);
A. V. Volotka {\it et al.},
Eur. Phys. J. D {\bf 33}, 23 (2005).
\bibitem{Sirlin1967}
A. Sirlin,
Phys. Rev. {\bf 164}, 1767 (1967);
A. Sirlin,
Nucl. Phys. B {\bf 71}, 29 (1974);
W. J. Marciano, A. Sirlin,
Nucl. Phys. B {\bf 88}, 86 (1975);A. Sirlin,
Rev. Mod. Phys. {\bf 50}, 573 (1978); 
Rev. Mod. Phys. {\bf 50}, 905 (1978);
A. Garc$\acute{\rm i}$a, M. Maya,
Phys. Rev. D {\bf 17}, 1367 (1978).
A. Sirlin,
Nucl. Phys. B {\bf 196}, 83 (1982);
W. J. Marciano, A. Sirlin,
Phys. Rev. Lett.  {\bf 56}, 22 (1986);
A. Czarnecki, W. J. Marciano, and A. Sirlin,
Phys. Rev. D {\bf 70}, 093006 (2004);
W. J. Marciano, A. Sirlin,
Phys. Rev. Lett.  {\bf 96}, 032002 (2006);
A. Czarnecki, W. J. Marciano, and A. Sirlin,
Phys. Rev. Lett. {\bf 99}, 032003 (2007).
\bibitem{Shann1971}
R. T. Shann, 
Nuovo Cimento A {\bf 5}, 591 (1971).
\bibitem{SUSY}
A. Krylov, M. J. Ramsey--Musolf,
Phys. Rev. Lett. {\bf 88}, 071804 (2002;);
S. Profimo, M. J. Ramsey--Musolf, and S. Tulin,
Phys. Rev. D {\bf 75}, 075017 (2007);
M. J. Ramsey--Musolf, S. Su,
Phys. Rep. {\bf 456}, 188 (2008).
\bibitem{Arzumanov2015} 
S. Arzumanov, L. Bondarenko, S. Chernyavsky,
  P. Geltenbort, V. Morozov, V. V. Nesvizhevsky, Yu. Panin, and
  A. Strepetov, Phys. Lett. B {\bf 745}, 79 (2015).
\bibitem{PDG2016} C. Partignani {\it et al.} (Particle Data Group),
  Chinese Physics C {\bf 40}, 100001 (2016).
\bibitem{Nico2006}
J. S. Nico {\it et al.}, Nature, {\bf 444}, 1059 (2006);
R. L. Cooper {\it et al.},
Phys. Rev. C {\bf 81}, 035503 (2010);
R. L. Cooper {\it et al.}, 
Nucl. Instrum. Meth. A {\bf 611}, 219 (2009);
M. J. Bales {\it et al.} (RDK II Collaboration),
Phys. Rev. Lett. {\bf 116}, 242501 (2016).
\bibitem{Ivanov2018} A. N. Ivanov, R. H\"ollwieser, N. I. Toitskaya,
  M. Wellenzohn, and Ya. A. Berdnikov, {\it Test of the Standard Model
    in neutron beta decay with polarized electron and unpolarized
    neutron and proton}, arXiv:1811.04853 [hep-ph].

\end{thebibliography}
\end{document}